\documentclass[reprint, aps, prfluids, onecolumn,  superscriptaddress, longbibliography, 12pt]{revtex4-2}

\usepackage{graphicx}
\usepackage{color}
\usepackage{amssymb}
\usepackage{amsmath}
\usepackage{tabularx}
\usepackage{bm}
\usepackage{hyperref}
\usepackage{ltxtable}
\usepackage{natbib}  
\usepackage{longtable}
\usepackage{parskip}
\usepackage{float}
\newcommand{\la}{\left\langle}
\newcommand{\ra}{\right\rangle}
\newcommand{\be}{\begin{equation}}
\newcommand{\ee}{\end{equation}}
\newcommand{\bse}{\begin{subequations}}
	\newcommand{\ese}{\end{subequations}}
\newcommand{\bea}{\begin{eqnarray}}
\newcommand{\eea}{\end{eqnarray}}
\newcommand{\ba}{\begin{array}}
	\newcommand{\ea}{\end{array}}

\usepackage[usenames,dvipsnames]{xcolor}
\hypersetup{
colorlinks,
citecolor=blue,
filecolor=blue,
linkcolor=blue,
urlcolor=blue}

\begin{document}

\title{Hydrodynamic Entropy and Emergence of Order in Two-dimensional Euler Turbulence}
\author{Mahendra K. Verma}
\email{mkv@iitk.ac.in}
\affiliation{Department of Physics, Indian Institute of Technology Kanpur, Kanpur 208016, India}
\author{Soumyadeep Chatterjee}
\email{inspire.soumya@gmail.com}
\affiliation{Department of Physics, Indian Institute of Technology Kanpur, Kanpur 208016, India}

\date{\today}

\begin{abstract}
{ Using numerical simulations, we show that the  asymptotic states of two-dimensional (2D) Euler turbulence exhibit large-scale flow structures due to nonzero energy transfers among small wavenumber modes. These asymptotic states, which depend on the initial conditions, are out of equilibrium, } and they are different from the predictions of Onsager and Kraichnan.   We propose ``hydrodynamic entropy'' to quantify order in 2D Euler turbulence; we show that this entropy decreases with time, even though the system is isolated with no dissipation and no contact with a heat bath. 
\end{abstract}
\maketitle

\section{Introduction}

Euler turbulence remains unsolved till date. In this paper, we address the energy flux and entropy of two-dimensional (2D) Euler turbulence. The equations of incompressible Euler flow are
\bea
\partial_t {\bf u} + {\bf u} \cdot \nabla {\bf u} = - \nabla p;~~~ \nabla \cdot {\bf u} = 0,
\label{eq:NS} 
\eea
where $ {\bf u}, p $ are the velocity and pressure fields respectively~\cite{Landau:book:Fluid,Frisch:book}. The above system is  \textit{isolated}, as it  lacks external force and viscous dissipation.  Consequently, the thermodynamic entropy of an Euler flow remains  constant~\cite{Landau:book:Fluid}.  However, when we  solve Euler equations with an ordered initial condition,
structurally, three-dimensional (3D) Euler turbulence becomes more random during its evolution~\cite{Cichowlas:PRL2005}, whereas  2D Euler turbulence tends to become more orderly~\cite{Fox:PF1973,Seyler:PF1975}.  Hence, the entropy of Euler turbulence needs a reexamination.

\citet{Onsagar:Nouvo1949_SH} modelled 2D Euler flow using a collection of point vortices interacting via logarithmic potential. Onsager  showed that for large energy, 2D Euler turbulence exhibits ``negative temperature" and a large  cluster of same-circulation vortices. Recently, \citet{Gauthier:Science2019} observed such giant vortices in an experiment involving 2D quantum fluid, thus providing an experimental  verification of Onsager's theory.  \citet{Billam:PRL2014} developed a first-principles realization of Onsager’s vortex model in a 2D superfluid. \citet{Miller:PRL1990} and \citet{Robert:JSP1991} extended Onsager's theory to continuum version of 2D Euler turbulence and computed entropy for the flow. { Using  tools of equilibrium statistical mechanics,  \citet{Bouchet:PRL2009}, and \citet{Bouchet:PR2012} derived multiple stationary states, namely a dipole and unidirectional flow (shear layer), for 2D Euler turbulence.  }  \citet{Pakter:PRL2018} provide a contrary viewpoint and showed that 2D Euler turbulence is out of equilibrium, and that a system of interacting vortices  becomes trapped in a nonequilibrium stationary state; these results deviate from the predictions of \citet{Onsagar:Nouvo1949_SH}. Refer to review articles by \citet{Eyink:RMP2006}, and \citet{Bouchet:PR2012} for an extensive discussion.

 \citet{Lee:QAM1952} and \citet{Kraichnan:JFM1973} provide an alternative framework for Euler turbulence. They showed that the  evolution of Fourier modes  of Euler equation follows Liouville’s theorem, and that the equilibrium solutions of Euler turbulence are 
   \bea
 E(k) = \frac{k^2}{\beta - \gamma k^2}~\mathrm{for~3D};
 \label{eq:Euler3D} \\
 E(k) = \frac{k}{\beta + \gamma k^2}~\mathrm{for~2D},\label{eq:Euler2D}
 \eea
 where $ \beta $ and $ \gamma $ are constants. Here, the Fourier modes form a microcanonical ensemble.  The derivation of Eqs.~(\ref{eq:Euler3D},\ref{eq:Euler2D})  involves two competing conservation laws: kinetic energy ($ \int d{\bf r} u^2/2 $) and kinetic helicity ($ \int d{\bf r} ({\bf u} \cdot \boldsymbol{\omega}) $) in 3D, and kinetic energy and enstrophy ($ \int d{\bf r} \omega^2/2 $) in 2D, where $ \boldsymbol{\omega} $ is the vorticity field. For some combinations of energy and enstrophy, 2D Euler turbulence yields $ \beta < 0$ or ``negative temperature"~\cite{Fox:PF1973,Kraichnan:ROPP1980}. \citet{Kraichnan:ROPP1980} provide a  detailed review of 2D Euler turbulence.
 
 
For a $ \delta $-correlated random velocity field as an initial condition, both 2D and 3D Euler turbulence follow the energy spectra of Eqs.~(\ref{eq:Euler2D}, \ref{eq:Euler3D}) with $ \gamma \approx 0 $~\cite{Verma:PTRSA2020,Verma:arxiv2020_equilibrium}. In addition, for an initial condition with large-scale structures, 3D Euler turbulence asymptotes to $ E(k) $ of  Eq.~(\ref{eq:Euler3D})~\cite{Cichowlas:PRL2005}. However, $ E(k) $ of 2D Euler turbulence differs from Eq.~(\ref{eq:Euler2D}) for coherent velocity field as an initial condition.  For example, for enstrophy-dominated 2D Euler turbulence, \citet{Fox:PF1973}  reported deviations from Eq.~(\ref{eq:Euler2D})   at  small wavenumbers. For parameters where $ \beta + \gamma k^2 \approx 0 $,  \citet{Seyler:PF1975} observed  large vortex structures, similar to those in a discrete vortex system~\cite{Joyce:JPP1973}.    \citet{Dritschel:JFM2015} studied the unsteady nature of 2D flow structures on a sphere. 	\citet{Robert:JFM1991}, and \citet{Bouchet:PR2012} analyzed such structures in the framework of equilibrium statistical mechanics.

The works of \citet{Fox:PF1973}, \citet{Seyler:PF1975},  \citet{Pakter:PRL2018}, \citet{Bouchet:PRL2009},  \citet{Dritschel:JFM2015}, and \citet{Modin:arxiv}  indicate that  2D Euler turbulence is out of equilibrium, contrary to the assumptions of \citet{Onsagar:Nouvo1949_SH} and \citet{Kraichnan:JFM1973}. \citet{Bouchet:PR2012} argued that even though nonequilibrium steady states of 2D Euler turbulence often break detailed balance, under weak force and zero viscosity, they may be described by microcanonical measures and entropy functional.  Bouchet and coworkers~\citep{Bouchet:PRL2009,Bouchet:PR2012,Kan:PTRSA2022}, and \citet{Modin:arxiv} explained the structures of  2D Euler turbulence in this framework. In this paper, we advance this theme by carefully examining the energy transfers and energy flux of 2D Euler turbulence. { We show that the small wavenumber modes exhibit nonzero energy transfers,} hence break the \textit{detailed balance},  which is a stringent criterion for equilibrium.  Thus, we demonstrate the nonequilibrium nature of 2D Euler turbulence. We also quantify the order of the structures using hydrodynamic entropy.

{ The outline of the paper is as follows.  In Sec.~\ref{sec:2d_euler}, we describe nonequilibrium nature of 2D Euler turbulence. We propose hydrodynamic entropy in Sec.~\ref{sec:hydro_entropy} to quantify this nature. We conclude the paper in Sec.~\ref{sec:conclusion}. }


\section{Nonequilibrium Nature of 2D Euler Turbulence}
\label{sec:2d_euler}

Prior to a detailed discussion on 2D Euler turbulence, we summarize the energy spectrum and flux of 3D Euler turbulence.  \citet{Cichowlas:PRL2005} simulated  3D  Euler turbulence with Taylor-Green vortex as an initial condition. For such simulations, in the early phase, the energy flows from large scales to small scales, and the energy flux is positive. After several eddy turnover times, the system approaches equilibrium with vanishing energy flux.   Refer to Appendix A for details.

For 2D Euler turbulence, we performed  pseudo-spectral simulations on a $ (2\pi)^2 $ box with a $ M^2 $ grid. Here, $ M=512 $.  We  dealiase  the code by setting  all the modes outside the sphere of radius $ M/3 $ to zero.  To conserve the total energy, we time evolve Eq.~(\ref{eq:NS}) using \textit{position-extended Forest-Ruth-like} (PEFRL) scheme~\cite{Omelyan:CPC2002,Forest:PD1990} with time step = $ 10^{-4} $.  We carried out three  runs with the following initial conditions:
\begin{enumerate}
	\item Run A: The initial velocity profile is taken as $ ( \sin 11x \cos 11y + \eta_x, - \cos 11x \sin 11y + \eta_y )  $, where $ (\eta_x, \eta_y) $ is random noise. We take $|\eta_x| \ll 1  $ and $|\eta_y| \ll 1$.
	
	\item Run B: The initial nonzero velocity Fourier modes are  $ {\bf u}(1,0) = (0,1)$,  $ {\bf u}(0,1) = (1,0)$, and  $ {\bf u}(1,1) = (-i,i)$.  
	
	\item Run C: The initial nonzero velocity Fourier modes are  $ {\bf u}(10,0) = (0,1)$,  $ {\bf u}(0,10) = (1,0)$, and  $ {\bf u}(10,10) = (-i,i)$.  
\end{enumerate}
\begin{figure}[hbtp]
	\includegraphics[width=\linewidth]{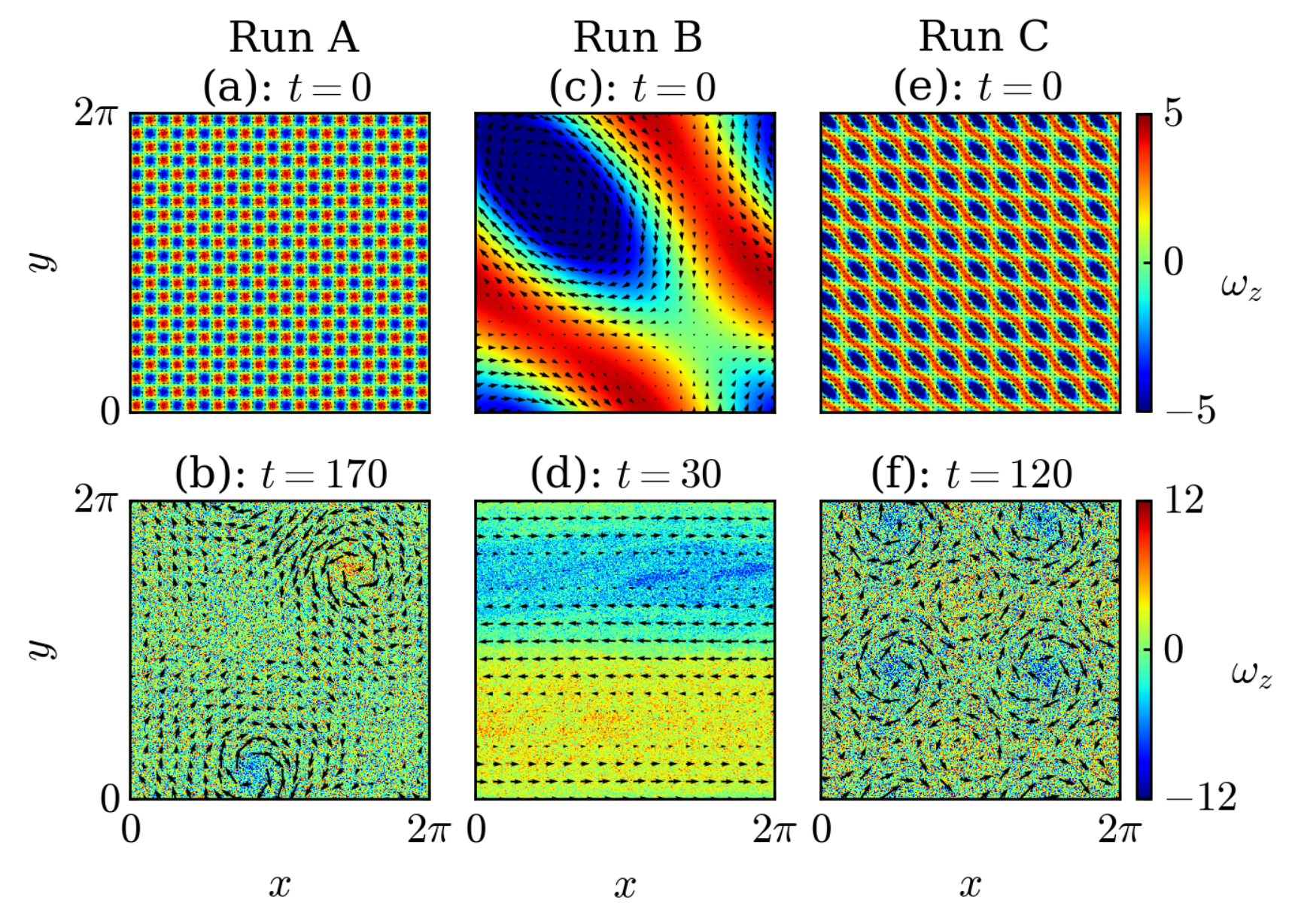}
	\caption{For Runs A, B, and C of 2D Euler turbulence: (a,c,e) the initial states, (b,d,f) the final states respectively. Here we plot the velocity field over the density plots of the vorticity field.} 
	\label{fig:2D_profile}
\end{figure}
We time advance Runs A, B, C up to 170, 30, and 120  turnover times  ($ 2\pi/U_\mathrm{rms} $) respectively.  { We observe Runs B and C reach the steady flow profiles after 10 to 20 eddy turnover times, whereas Run A reaches the steady state after 100 eddy turnover times.   We label these states as \textit{asymptotic states}. In this paper, we show that these states are out of equilibrium.  The dependence of the asymptotic states on initial condition is consistent with earlier works, e.g., \citet{Celani:PRL2010}.}  The top and bottom panels of Fig.~\ref{fig:2D_profile} illustrate respectively the initial  and asymptotic  states of the three runs.  Here, the velocity field is superposed over the density plots of the vorticity field.   { Runs A, B, C asymptote respectively to a vortex-antivortex pair~\cite{Onsagar:Nouvo1949_SH},  a unidirectional flow (shear layer), and  four vortex-antivortex pairs.  The above  large-scale flow structures are embedded in small-scale noisy flow.  Similar structures have been observed  by  \citet{Bouchet:PRL2009}, and \citet{Dritschel:JFM2015}.}

\begin{figure}[hbtp]
	\includegraphics[width=0.8\linewidth]{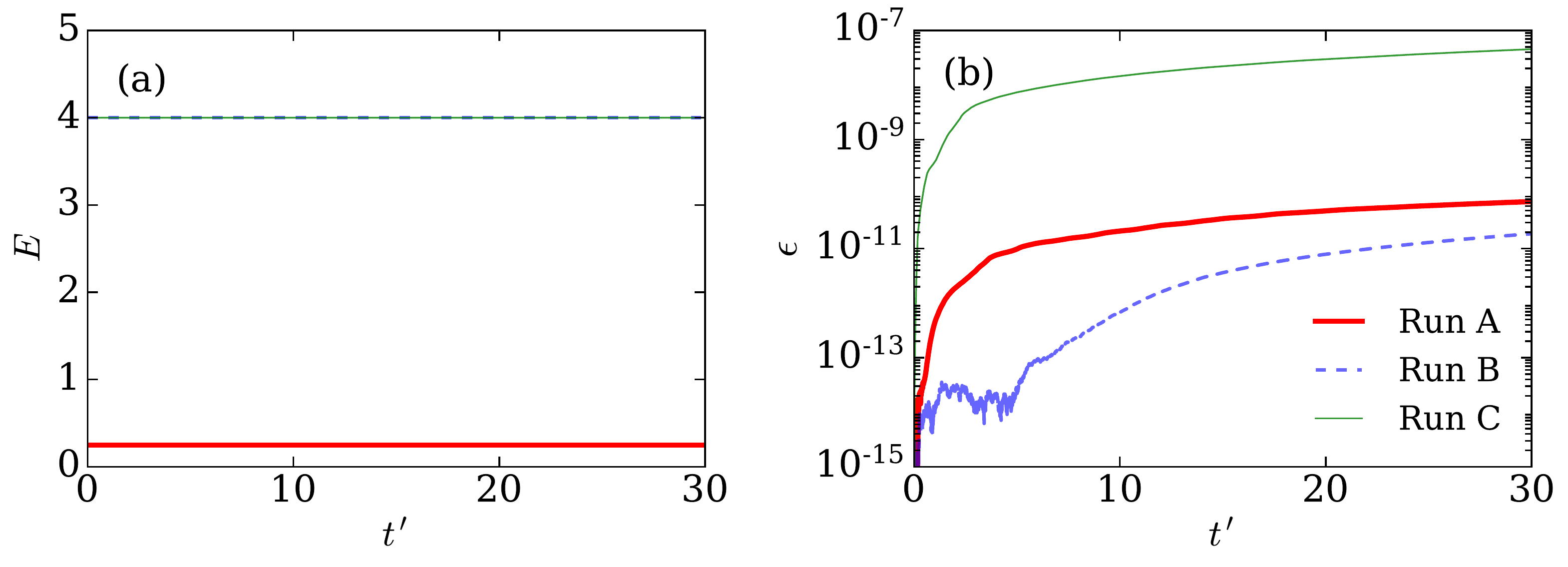}
	\caption{{\color{blue}(a) For Runs A, B, C of 2D Euler turbulence, (a) time series of the total energy, $E$, and (b) its relative error, $\epsilon$. We employ normalized time $ t'=3t/17, t, t/4 $ for the three runs. }}
		  	\label{fig:E_u}
\end{figure}

\begin{figure}[hbtp]
	\includegraphics[width=0.8\linewidth]{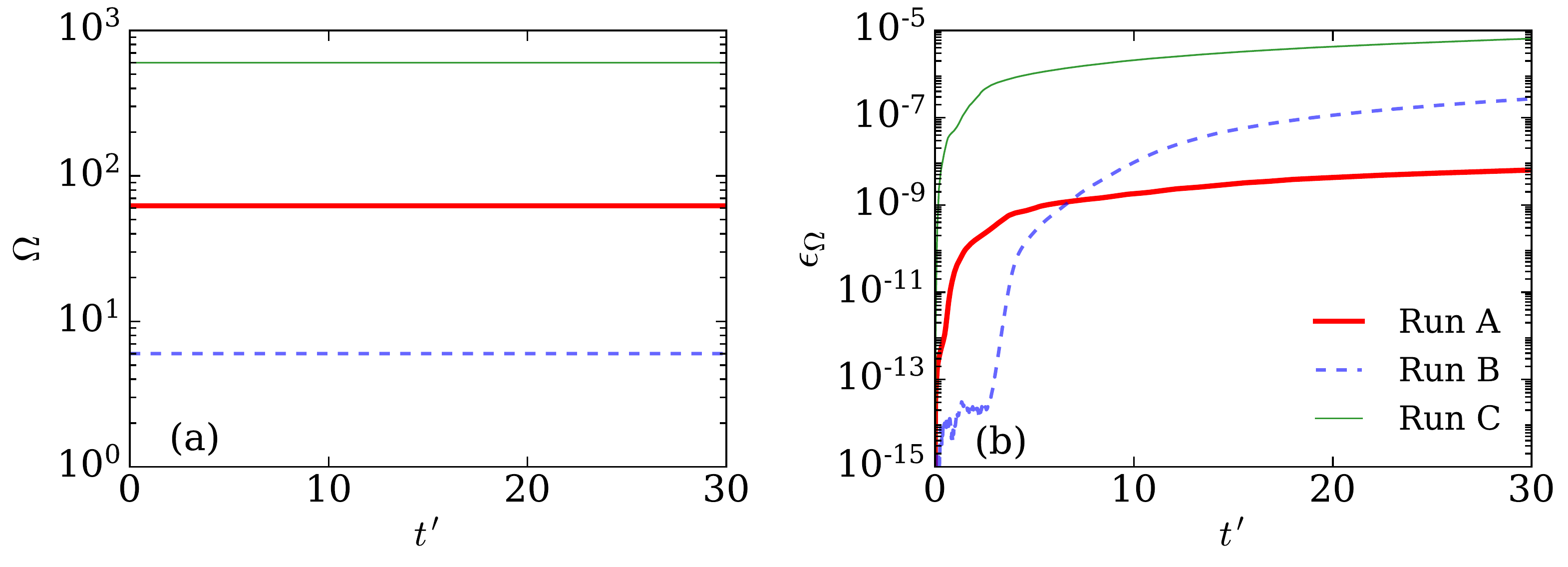}
	\caption{{ For Runs A, B, C of 2D Euler turbulence, (a) time series of  total enstrophy, $\Omega$, and (b) its relative error, $\epsilon_{\Omega}$, as a function of $t'$, where $ t'=3t/17, t, t/4 $ for the three runs. }}	\label{fig:E_omega}
\end{figure}

For Runs A, B, and C, total energy $ E =  \int d\textbf{r} (u^2/2)/\int d\textbf{r} = 0.2500954, 4, 4 $, while total enstrophy $ \Omega = \int d\textbf{r} (\omega^2/2)/\int d\textbf{r} = 62.17, 6, 600 $, respectively; and these quantities are conserved.  { For validation of our numerical codes, we plot the time series of the total energy and total enstrophy, along with their relative errors in Figs.~\ref{fig:E_u} and ~\ref{fig:E_omega}. The relative errors for energy and enstrophy are defined as
 	\be
 	\epsilon(t) = \frac{|E(t)-E(t=0)|}{E(t=0)};~~~~
 	\epsilon_\Omega(t) = \frac{|\Omega(t)-\Omega(t=0)|}{\Omega(t=0)}.
 	\ee
 Since the three runs take different amounts of time to reach their respective asymptotic states, in the plots, we employ normalized time, $ t' =  3t/17, t, t/4 $, for the Runs A, B, and C respectively.	For the asymptotic states of Runs A, B, and C, the relative errors in energy ($\epsilon$) are $ 7.3\times 10^{-11} $, $ 1.9 \times 10^{-11} $ and $ 4.5 \times 10^{-8} $ respectively, while those in  entropy ($\epsilon_{\Omega}$) are  $ 6.3\times 10^{-9} $, $ 2.7\times 10^{-7} $,  and $ 6.5\times 10^{-6} $ respectively.  } 

\begin{figure}[hbtp]
	\includegraphics[width=0.9\linewidth]{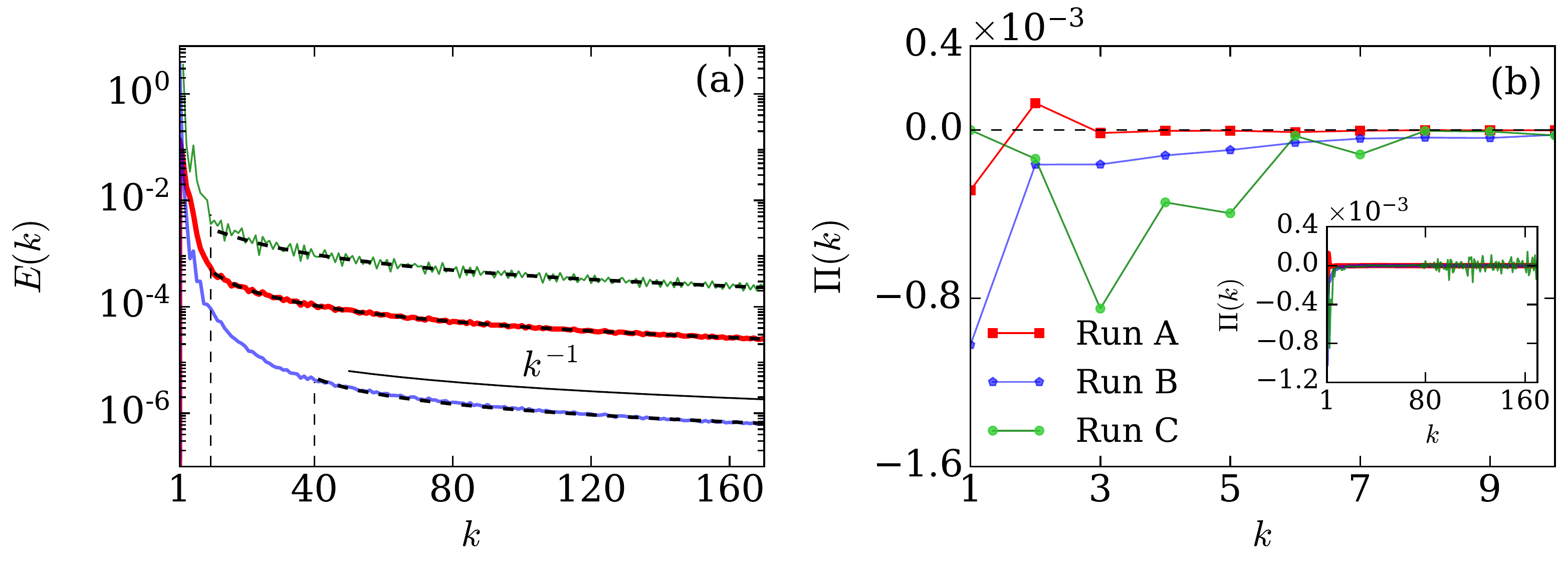}
	\caption{For the  Runs A, B, C of 2D Euler turbulence: plots of the averaged energy spectra, $ E(k)  $, and fluxes, $ \Pi(k) $, of the asymptotic states.  In (a), the best-fit curves of Eqs.~(\ref{eq:RunA}-\ref{eq:RunC}) are shown as dashed curves. Figure (b) exhibits $ \Pi(k) $ for small $ k $'s, while inset shows $ \Pi(k) $ for the whole range.}	\label{fig:2D_Ek}
\end{figure}

{ In Fig.~\ref{fig:2D_Ek}(a,b), we plot the averaged energy spectra and fluxes of the three runs. Here, we average 6500, 1000, and 10000 frames of Runs A, B, and C in the respective time intervals $(105, 170), (20, 30)$ and $(20, 120)$.} For intermediate and large wavenumbers, $ E(k) $ of the three runs follow Kraichnan's predictions  [Eq.~(\ref{eq:Euler2D})]:
\bea
\mathrm{Run~A:} & ~ & E(k) = \frac{k}{-3230+237 k^2}~	\mathrm{for}~k > 10,  \label{eq:RunA} \\
\mathrm{Run~B:}  & ~ & E(k) = \frac{k}{-6357840+9361 k^2}~\mathrm{for}~k > 40,  \label{eq:RunB}\\
\mathrm{Run~C:}  & ~ & E(k) =\frac{k}{888+26 k^2}~\mathrm{for}~k > 10, \label{eq:RunC}
	\eea
with constants having significant errors. {  We construct these functions using \textit{non-linear least squares fit} of the numerical data of $k/E(k)$ to $\beta + \gamma k^{2}$. We employ the standard \textit{scipy.optimize.curve\_fit} function of Python for the computation.} 

\begin{figure}[hbtp]
	\includegraphics[width=0.5\linewidth]{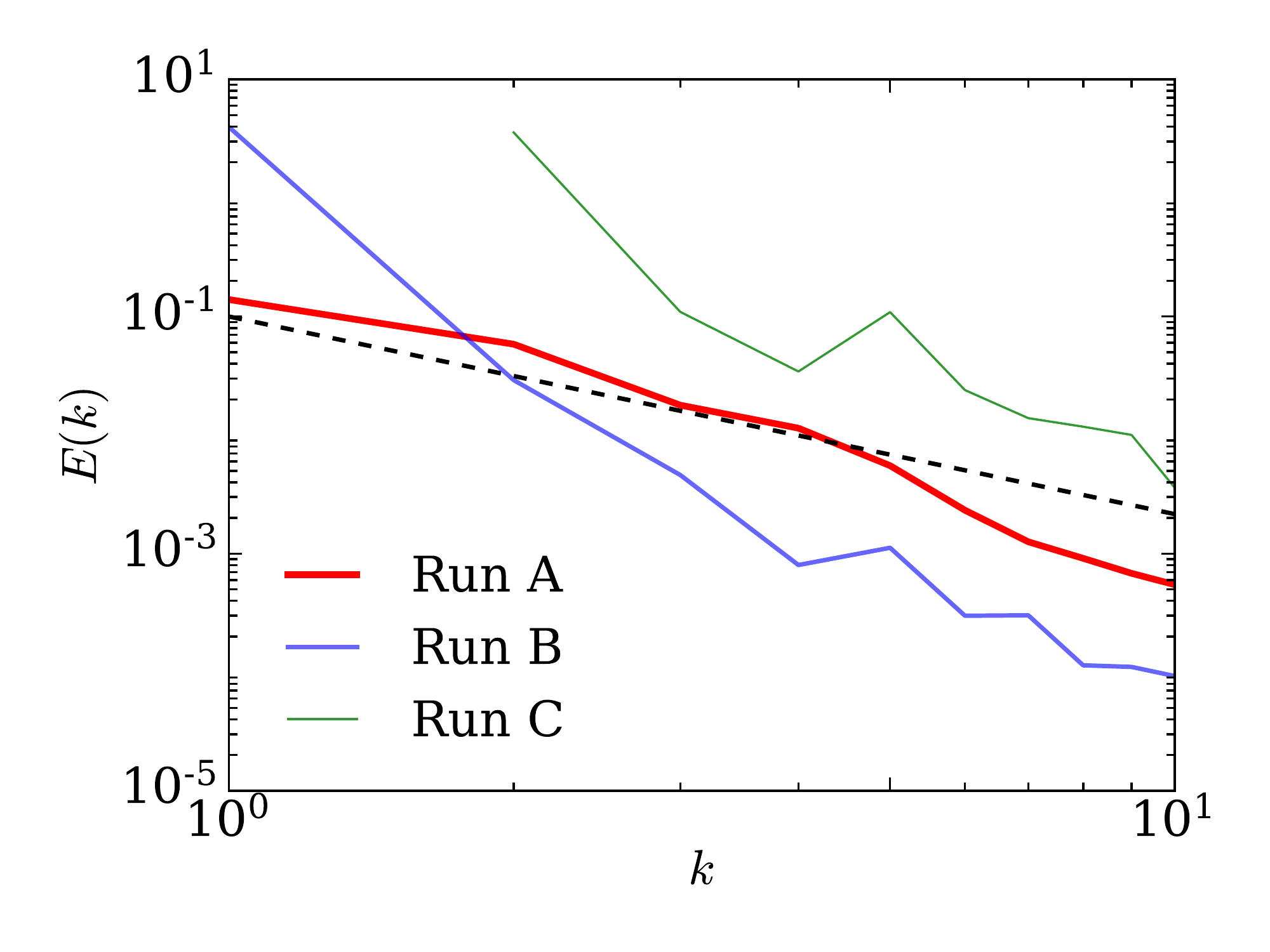}
	\caption{ For the  Runs A, B, C of 2D Euler turbulence, plots of the averaged energy spectra for small wavenumbers  $k\in[1, 10]$. The black-dashed line represents $ k^{-5/3} $ spectrum. Clearly,    Runs B and C do not exhibit  $ k^{-5/3} $ spectrum, while Run A  matches with $ k^{-5/3} $ for very small region.}
	\label{fig:2D_Ek_sm_k}
\end{figure}

{ Based on Eqs.~(\ref{eq:RunA}-\ref{eq:RunC})  and 
Figure~\ref{fig:2D_Ek}(a), we claim that the large wavenumber modes are in equilibrium.   \citet{Fjortoft:Tellus1953} and \citet{Nazarenko:book:WT} showed that for 2D hydrodynamic turbulence, $ k_E \le \sqrt{\Omega/E} \le k_\Omega $, where $ k_E $ and $ k_\Omega $ are the centroids of energy and enstrophy respectively. For Runs A, B, and C, $ \sqrt{\Omega/E} = 15.8, 1.2,  12.2$. Hence,  the  wavenumbers far beyond  $  k_\Omega $ are dominated by enstrophy.  In this regime, our simulations reveal that $ E(k) \propto k^{-1} $, which corresponds to an equipartition of enstrophy. That is, modal enstrophy, $ E_\omega({\bf k}) =E({\bf k}) k^2 = \mathrm{constant} $, which leads to $ E(k) \sim 2\pi k/k^2 \sim k^{-1} $. Moreover, Runs A and B exhibit $ \beta < 0 $ or ``negative temperature" \cite{Onsagar:Nouvo1949_SH,Kraichnan:JFM1973,Kraichnan:ROPP1980}, which is related to the emergence of large scale structures.  Note that  \citet{Onsagar:Nouvo1949_SH} and \citet{Kraichnan:JFM1973} assumed the 2D Euler flow to be in equilibrium.

Equations~(\ref{eq:RunA}-\ref{eq:RunC})  and 
Figure~\ref{fig:2D_Ek}(a) reveal that the small wavenumber modes  deviate strongly from Kraichnan's predictions for equilibrium Euler turbulence, consistent with the works of  \citet{Fox:PF1973}.  In the following discussion,  we will show that the small-wavenumber modes of 2D Euler are out of Equilibrium  from various perspectives. As illustrated in Fig.~\ref{fig:2D_Ek_sm_k}, for small wavenumbers, $ E(k) $'s   do not follow $ k^{-5/3} $ energy spectrum, except for a small range of wavenumbers for Run A.  This is expected because Euler turbulence is very different from 2D hydrodynamic turbulence, which exhibits $ E(k) \propto k^{-5/3} $ when  the intermediate scales are forced~\cite{Kraichnan:JFM1971_2D3D}.
} 

\begin{figure}[hbtp]
\includegraphics[width=\linewidth]{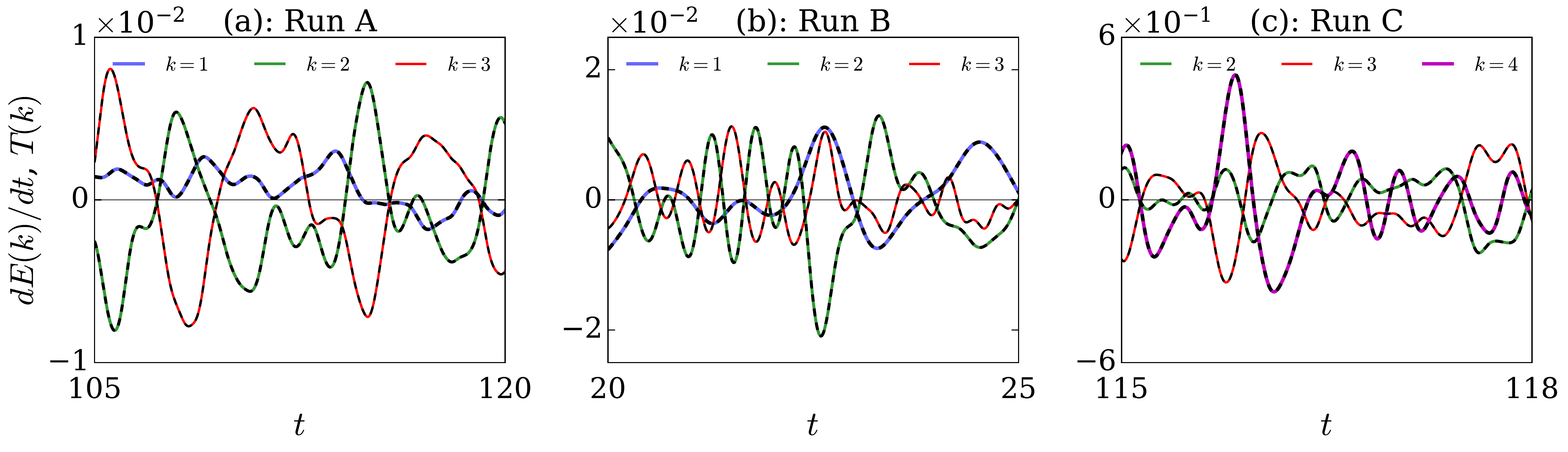}
\caption{{ For Runs A, B, C, plots of the time evolutions of $dE(k)/dt$ (solid lines) and $T(k)$ (dashed lines) for three different wavenumbers.} }\label{fig:Tk_time}
\end{figure}

{ We complement the spectral studies with a  quantification of energy transfers and fluxes. Euler turbulence lacks external forcing and dissipation, hence the temporal evolution of $ E(k) $ is given by~\cite{Frisch:book,Majda:PNAS2015,Verma:book:ET,Verma:PRE2022,	Verma:JPA2022}
\be
\frac{dE(k,t)}{dt} = T(k,t),
\label{eq:Tk}
\ee 
where $ T(k,t) $ is the nonlinear energy transfer to wavenumber shell $ k $.  If Euler turbulence were to be in equilibrium, then for all $ k $'s,  $ dE(k,t)/dt=0 $, implying that $ T(k,t) =0 $.  Consequently, the energy flux, $ \Pi(k,t) = -\int_0^k T(k',t) dk' $ would vanish for all wavenumber spheres. This test is an alternative one to the entropy maximization principle~\cite{Verma:PRE2022}. Energy conserving systems, as well as Hamiltonian systems, are expected to approach equilibrium asymptotically~\cite{Landau:book:StatMech,Majda:PNAS2015}. It has been shown that 3D Euler turbulence asymptotically reaches the equilibrium state with zero energy flux~\cite{Cichowlas:PRL2005}. But,  we show below that $ T(k,t) $ and $ \Pi(k,t) $ for 2D Euler turbulence are nonzero for small $ k $'s, thus 2D Euler turbulence is out of equilibrium.

For 2D Euler turbulence, we illustrate $ T(k,t) $ for the dominant wavenumber shells ($ k $ = 1 to 4) in Fig.~\ref{fig:Tk_time}.  Clearly, these $ T(k,t) $'s  fluctuate significantly, and $ dE(k,t)/dt \approx T(k,t) $, thus validating Eq.~(\ref{eq:Tk}). For small $ k $'s, the nonzero $ T(k,t) $'s yield negative energy flux.  In particular, $ \min [\la \Pi(k)  \ra] \approx -3 \times10^{-4}, -10^{-3}, -8.4\times 10^{-4}$ for Runs A, B, C respectively. Note that the finite $ T(k,t)$ and $\Pi(k,t)$ for 2D turbulence are much larger than the corresponding quantities of 3D Euler turbulence (see Appendix A).
The nonzero $ T(k,t)  $ and $ \Pi(k,t) $ break the \textit{detailed balance of energy transfers} and indicate nonequilibrium nature of 2D Euler turbulence.}

Based on the above observations, we conclude that the intermediate and large wavenumber shells are in equilibrium, but the small wavenumber shells are out of equilibrium. However, there is an exception to the above rule. For $ \delta $-correlated velocity field as an initial condition, 2D Euler turbulence exhibits equilibrium solution with $ E(k) \propto k $~\cite{Verma:arxiv2020_equilibrium}. For this particular case, $ \gamma \approx 0 $, and $ k_\Omega $ exceeds the grid size. 

In classical  literature of  hydrodynamics,	the thermodynamic  entropy of Euler turbulence is taken to be constant~\cite{Landau:book:Fluid}. However, as discussed above, the disorder in Euler turbulence varies with time. Hence, the thermodynamic entropy cannot capture the disorder in Euler turbulence.  In Section~\ref{sec:hydro_entropy}, we define ``hydrodynamic entropy" that can describe the disorder in Euler turbulence.

\section{Hydrodynamic Entropy}
\label{sec:hydro_entropy}

\begin{figure}[hbtp]
	\includegraphics[width=0.7\linewidth]{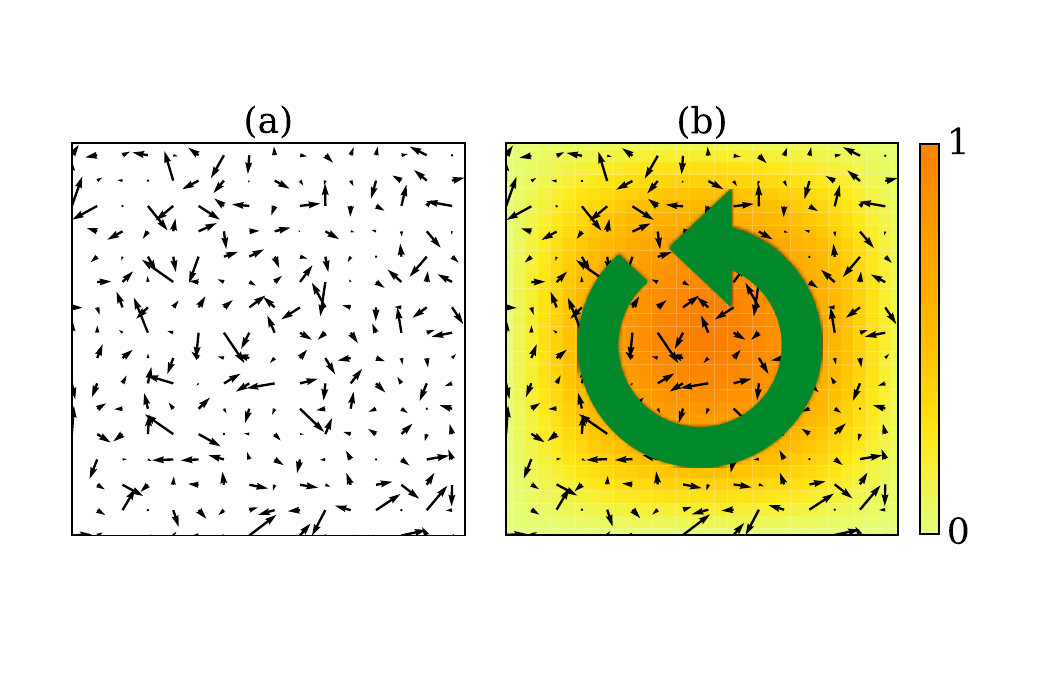}
	\caption{(a) In a 2D gas in thermal equilibrium, the molecules move randomly with the sound speed $ C_s $.  (b) A schematic diagram illustrating the velocities of  molecules in a hydrodynamic vortex. 
	}
	\label{fig:entropy_hydro}
	
\end{figure}

Consider two gaseous systems  shown in Fig.~\ref{fig:entropy_hydro}. In Fig.~\ref{fig:entropy_hydro}(a), the molecules are in thermal equilibrium, and they move randomly with thermal speed $ C_s $. Thermodynamic entropy provides a good measure for the disorder in such a system. 

In Fig.~\ref{fig:entropy_hydro}(b), the molecules co-move in a hydrodynamic vortex.   This system is out of equilibrium because 2D hydrodynamic vortices coalesce to form large vortices via inverse cascade of energy.   To quantify disorder in an hydrodynamic flow, we scale separate  the microscopic thermal  processes and  the macroscopic fluid processes, and define two different entropies for them~\cite{Verma:EPJB2019,Verma:PTRSA2020}.  Thermodynamic entropy  is employed for the  microscopic processes~\cite{Kundu:book,Choudhuri:book:Astro}, whereas ``hydrodynamic entropy", to be defined below, for describing macroscopic order. For an Euler flow, only hydrodynamic entropy would be meaningful because thermodynamic entropy is zero for this case.

We  employ \textit{Shannon entropy}~\cite{Shannon:BELL1948} to quantify the disorder of the fluid structures at macroscopic level.  For the same, we postulate that  the probability of occurrence of a Fourier mode with wavenumber $ {\bf k} $ is $ p_{\bf k} = E({\bf k})/E$, where $ E({\bf k}) $ is the modal energy, and  $ E $ is the total energy. Now, the hydrodynamic entropy of the flow is defined as 
\be
S = -\sum_{\bf k}   p_{\bf k}  \log_2 ( p_{\bf k} ).
\label{eq:entropy}
\ee 
The above entropy, which is defined for a snapshot, can be applied to any fluid flow. Note that  hydrodynamic entropy is very different from  thermodynamic entropy, which depends on temperature and volume of the system.  At present, it is not apparent if hydrodynamic entropy and thermodynamic entropy could be put on a similar footing. It would be interesting to explore whether we can relate the two in a  way  \citet{Landauer:IBM2000} connected computation to entropy.  

It is important to keep in mind that Shannon entropy [Eq.~(\ref{eq:entropy})] has been used earlier to quantify order in images, music, messages, DNA, as as well as in   variety of  flow phenomena, e.g., boundary layers, transitions, etc.~\cite{Aubry:JSP1991}.  There are many other measures of entropy.  For example, \citet{Clark:PRF2020} employed Kolmogorov-Sinai entropy for quantifying 2D hydrodynamic flow, whereas \citet{Drivas:Arxiv2022} quantified entropy of Euler flow in terms of accessible phase space.

Euler flow is an ideal case where thermal processes are ignored (due to zero viscosity). Therefore, Euler flow has only hydrodynamic entropy. The entropies of the vortex solutions of \citet{Onsagar:Nouvo1949_SH} and  \citet{Miller:PRL1990}  are hydrodynamic, however, these measures may differ from those computed using Eq.~(\ref{eq:entropy}). Interestingly, Kraichnan's equilibrium solution with $ \gamma=0 $ corresponds to $ \delta $--correlated  velocity field~\cite{Lee:QAM1952,Kraichnan:JFM1973,Verma:PTRSA2020}. For such a flow, $ p_{\bf k} = 1/M$ and the hydrodynamic entropy $ S = \log_2(M) $, where $ M $ is the number of modes of the system; this is the maximum possible hydrodynamic entropy for a flow with $ M$ degrees of freedom.

\begin{figure}[htbp]
 	\includegraphics[width=0.55\linewidth]{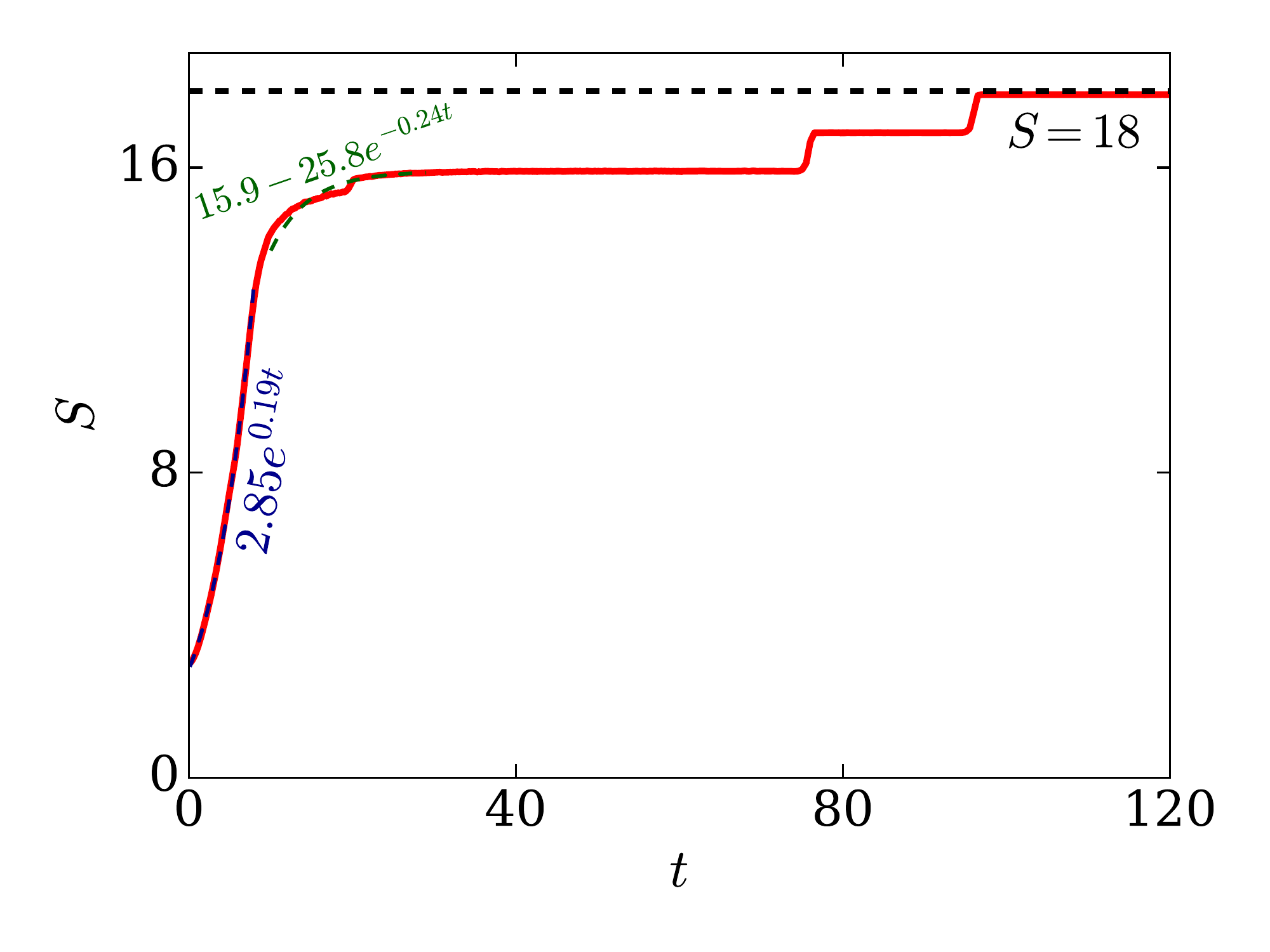}
 	\caption{In 3D Euler turbulence, hydrodynamic entropy exhibits monotonic growth, with  exponential increase in the beginning and saturation in the end.}
 	\label{fig:3D_entropy}
\end{figure}

Before reporting the hydrodynamic entropy  of 2D Euler turbulence, we describe the hydrodynamic entropy of 3D Euler turbulence.  For the 3D Euler flow described in Appendix A, we  compute the entropy using  Eq.~(\ref{eq:entropy}) and plot its time series in Fig.~\ref{fig:3D_entropy}. We derive the functional dependences of $ S(t) $ using the non-linear least squares method.   In the initial phase, the hydrodynamic entropy increases exponentially,  after which it approaches  the maximum possible value $ S = \log_2(M)  = 18.3$, where $ M = (4\pi/3)(128/3)^3$ is the number of modes of the system.  We observe that for 3D Euler turbulence,  the approach to equilibrium  is slow.   Since $ S $ increases monotonically in time, we claim that 3D Euler turbulence evolves from order to disorder.

\begin{figure}[hbtp]
	\includegraphics[width=0.6\linewidth]{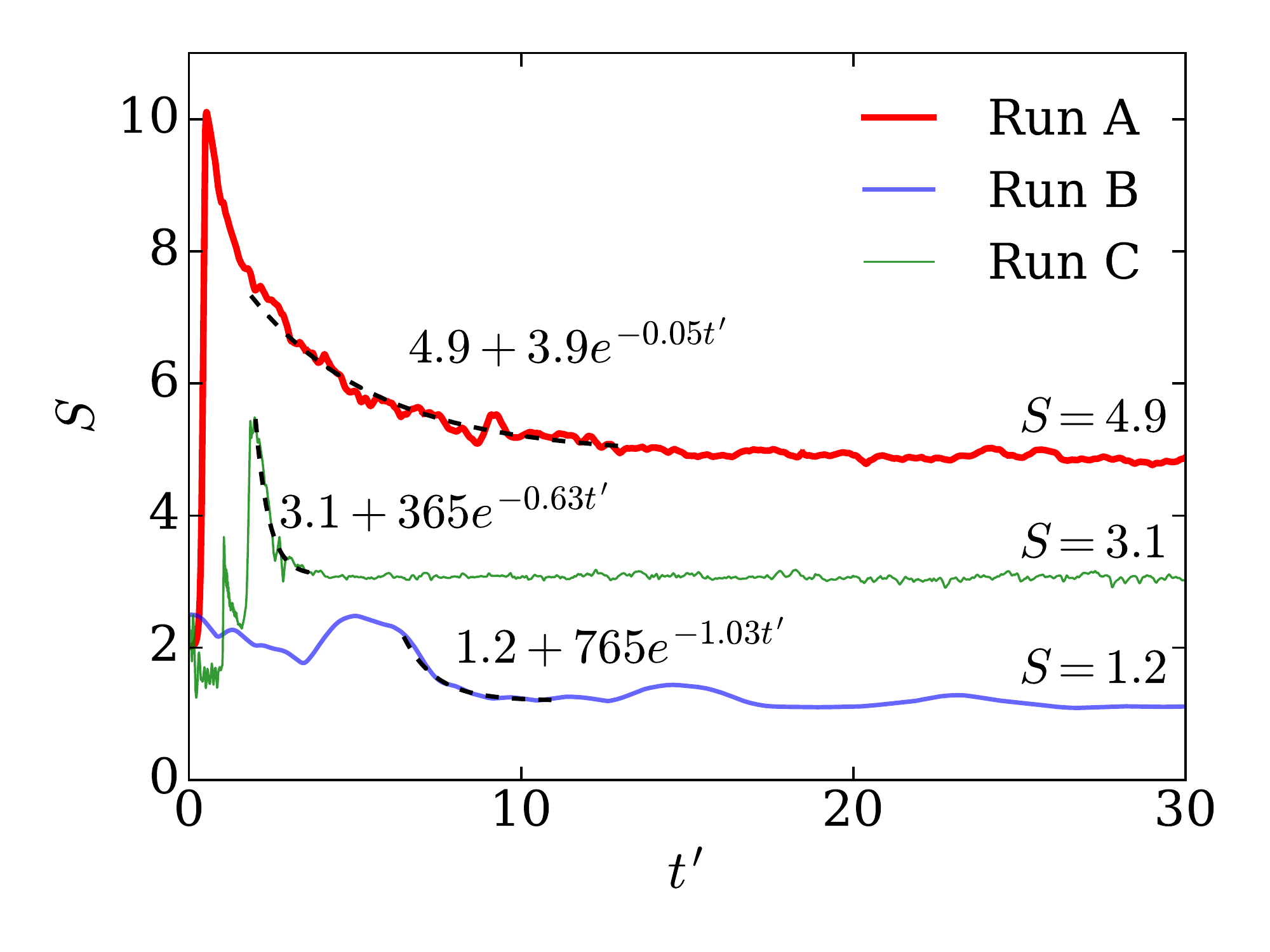}
	\caption{For Runs A, B, C of 2D Euler turbulence, the temporal evolution of hydrodynamic entropies with $ t'=3t/17, t, t/4 $ respectively.
		In each case, after initial transients, the entropy decreases with time and asymptotes to an approximate constant value. } 
	\label{fig:2D_entropy}
\end{figure}

Now, we compute the hydrodynamic  entropies of the  2D Euler flows of Runs A, B, and C, and plot  the entropy time series in Fig.~\ref{fig:2D_entropy}.  { The duration and time scales of the three runs are quite different. Hence, for a proper comparison, we normalize the time appropriately, i.e., $ t'=3t/17, t, t/4 $ for Runs A, B, and C respectively.} We observe that for each case, the entropy fluctuates in the early stages, after which it decreases exponentially  to  an asymptotic value. { Here too, we  employ the non-linear least squares method  to derive the functions that describe the decrease in $S$.}  The asymptotic entropies for Runs A, B, and C are 4.9, 1.2, and 3.1 respectively. These values are smaller than the maximum possible value, which is $ \log_2(M) \approx 16.5$, where $ M = \pi (512/3)^2 $ is the number of active dealiased  modes.  Note, however,  that $ S  $ for all the runs exhibit small fluctuations in the asymptotic regime due to the dynamic nature of the flow. {  Recall the temporal fluctuations of $ T(k,t) $ and $ E(k,t)$ described in Section~\ref{sec:2d_euler}. It is also important to note that the evolution of  hydrodynamic entropy for 2D Euler turbulence is very different that of 3D Euler turbulence.}

\begin{figure}[hbtp]
		\includegraphics[width=0.56\linewidth]{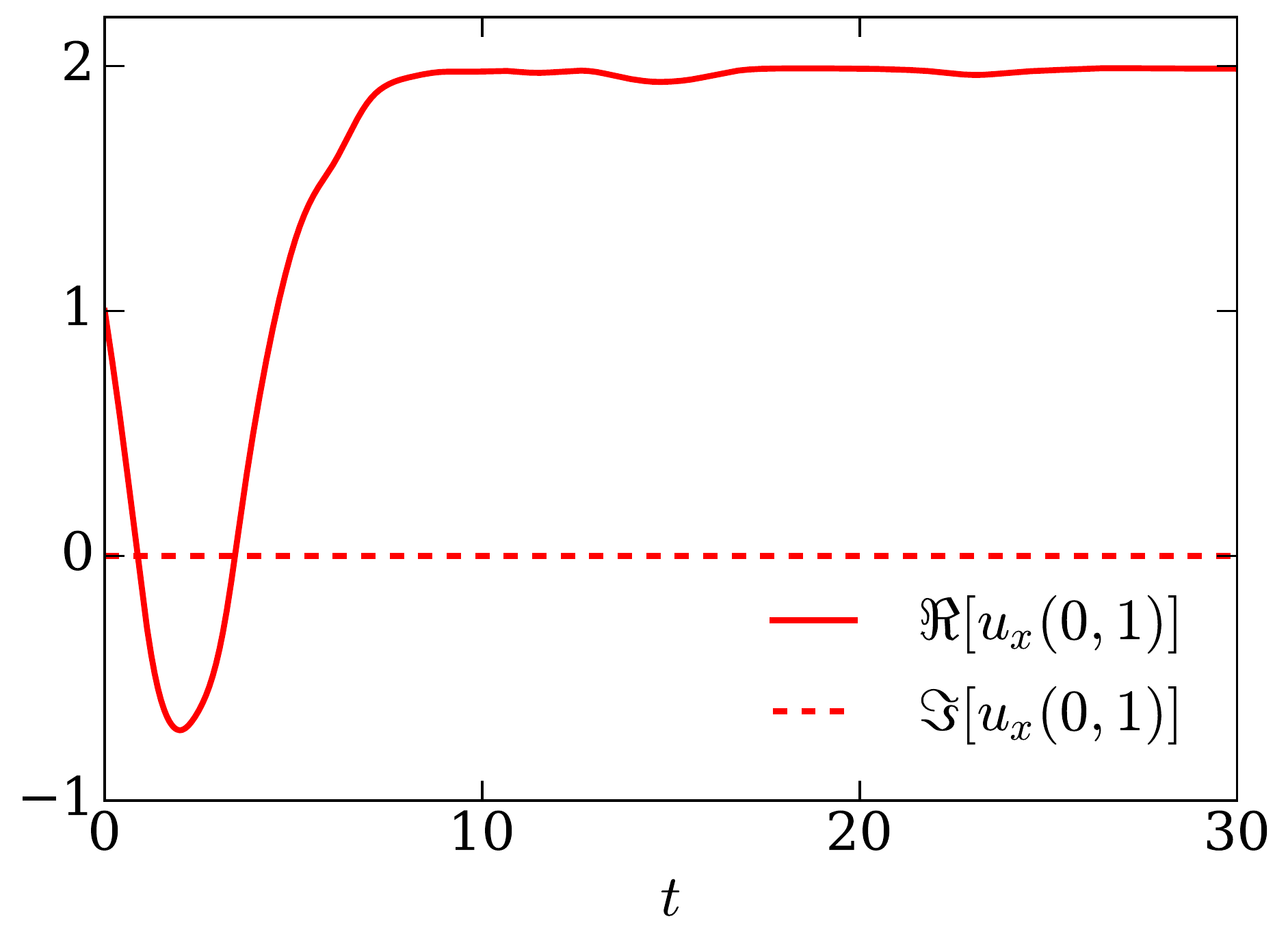}
		\caption{For  Run B of 2D Euler turbulence, time series of  $ \Re[{u_x(0,1) }]$ and  $ \Im[{u_x(0,1) }]$. In this run, $ {\bf u}(0,1) $ is the most dominant Fourier mode. }
		\label{fig:u_0_1}
	\end{figure}

{ The asymptotic state of Run B is the most ordered one (having least entropy) among the three runs.} In the asymptotic regime of  Run B, the Fourier modes $ {\bf u}(0,\pm 1) $ contains 99\% of the total energy, hence, their probabilities are approximately 1/2 each. Therefore, these two modes yield $ S \approx \log_2(2) = 1$, while the other modes contribute the rest  (0.2). {  To quantify the evolution of these modes, in Fig.~\ref{fig:u_0_1},  we plot the time series of $ \Re[{u_x(0,1) }]$ and  $ \Im[{u_x(0,1) }]$ of Run B. As shown in the figure, $ \Im[{u_x(0,1) }] = 0$, while $ \Re[{u_x(0,1) }]$ approaches nearly a constant value for $ t>10 $.  The fluctuations in $ u_x(0,1)  $ is consistent with those in $ T(k,t) $ discussed in Section~\ref{sec:2d_euler}.}

In contrast, for Run C, the  seven most energetic modes  yield an approximate entropy of 2.3, while the rest of them contribute the remaining 0.8.  The Fourier modes of Run A  are more wide-spread than those of Runs B and C, hence Run A has the maximum entropy among  the three runs. { Note that the shear layer of Run B has the least entropy, but the vortex-antivortex pair of Run A has the maximum entropy, which is still smaller than the maximum possible value of 16.5.}

Thus, we show that the hydrodynamic entropy of 2D Euler turbulence decreases  with time for a significant duration, even though the system is  isolated.  Thus, 2D Euler turbulence is a rare isolated  system that exhibits evolution from \textit{disorder to order}.     

\section{Discussions and Conclusions} 
\label{sec:conclusion}
 
\citet{Kraichnan:JFM1973} and \citet{Onsagar:Nouvo1949_SH} assumed that 2D Euler turbulence reaches equilibrium.  Kraichnan argued that 2D Euler turbulence exhibits the energy spectrum of Eq.~(\ref{eq:Euler2D}), whereas  \citet{Onsagar:Nouvo1949_SH} advocated a large cluster of same-circulation vortices.  We performed numerical simulations to test whether 2D Euler turbulence is in equilibrium or out of equilibrium.

Our numerical simulations, as well as several past ones, report  that at small wavenumbers, the energy spectrum and flux of 2D Euler turbulence differ from those predicted by \citet{Kraichnan:JFM1973}.  These differences arise due to the nonequilibrium nature of 2D Euler turbulence. { It is important to note that 2D Euler turbulence is one of the few energy-conserving systems that do not thermalize (or approach equilibrium).} Note, however, that several experiments on  quantum fluids report consistency with Onsager's theory~\cite{Gauthier:Science2019}. Hence, the relationship between Euler turbulence and quantum fluids needs to be examined carefully. We also remark that the flow structures of 2D Euler turbulence do not exhibit long-range order, as in a ferromagnet. Instead, the flow structures resemble vortex-antivortex pairs of Berezinskii–Kosterlitz–Thouless transition~\cite{Chaikin:book}.

More importantly, we propose hydrodynamic entropy to quantify disorder in Euler turbulence. We observe that the hydrodynamic entropy of 3D Euler turbulence increases monotonically with time, whereas it decreases for 2D Euler turbulence.  Thus, 2D Euler turbulence is a unique \textit{isolated} system that exhibits evolution form disorder to order. This feature arises due to the inverse energy cascade, which is a property of 2D hydrodynamics~\cite{Kraichnan:JFM1971_2D3D,Lesieur:book:Turbulence,Frisch:book,Verma:book:ET}.  Hence, the emergence of hydrodynamic order in 2D Euler turbulence has a dynamic origin.  Note that the thermodynamic entropy  of Euler turbulence remains constant throughout its evolution. Therefore, \textit{the decrease in hydrodynamic entropy with time does not violate second law of thermodynamics.}

{ Euler equation is time reversible due to an absence of viscous dissipation~\cite{Frisch:book}. This is the reason why  thermodynamic entropy of Euler turbulence is constant~\cite{Landau:book:Fluid}. Note, however, that  the solutions of Euler equation exhibits irreversibility due to their chaotic and nonequilibrium nature.  As we show in this paper, 2D Euler turbulence  is a special energy-conserving system that exhibit nonequilibrium behaviour.  However,  3D Euler turbulence reaches equilibrium asymptotically where detailed balance is preserved statistically~\cite{Cichowlas:PRL2005}.  Interestingly, the hydrodynamic entropy captures the irreversibility and disorder of 2D and 3D Euler turbulence quite well. 

We conclude this paper by emphasizing that  thermal and hydrodynamic processes, with or without viscosity, are multiscale phenomena where hierarchical energy transfers play a critical role~\cite{Frisch:book,Verma:JPA2022}.  Recently, \citet{Verma:EPJB2019,Verma:PTRSA2020} attributed irreversibility in a turbulent flow to the asymmetric energy transfers (e.g., forward cascade in 3D Navier-Stokes equation), which is a hydrodynamic property.   Following a similar approach, in this paper, we propose hydrodynamic entropy that successfully  captures the evolution of 2D Euler turbulence from disorder to order.}

\section*{Acknowledgements}
The {authors thank} Arul Lakshminarayan, Siva Chandran, Shashwat Bhattacharya, Pankaj Mishra, Sagar Chakraborty, and Anurag Gupta for useful  discussions.   This work is supported by the project 6104-1  from the Indo-French Centre for the Promotion of Advanced Research (IFCPAR/CEFIPRA).  Soumyadeep Chatterjee is supported by INSPIRE fellowship (IF180094) from Department of Science \& Technology, India.

\appendix
\section{Evolution of 3D Euler Turbulence}
\label{sec:3d_euler}

{ In this Appendix, we summarize the energy spectrum and flux of 3D Euler turbulence.  \citet{Cichowlas:PRL2005} simulated  3D  Euler turbulence with Taylor-Green vortex as an initial condition. For such simulations, in the early phase, the energy flows from large scales to small scales, and the energy flux is positive. After several eddy turnover times, the system approaches equilibrium with vanishing energy flux and $ E(k)  $ given by Eq.~(\ref{eq:Euler3D}). 
	
	To compute the hydrodynamic entropy for 3D Euler turbulence, we simulated 3D Euler flow.	We performed our simulation on a $ (2\pi)^3 $ box with a $ M^3 $ grid. Here, $ M=128 $.  We  dealiase  the code by setting  all the modes outside the sphere of radius $ M/3 $ to zero.  We time evolve Eq.~(1) of the main text using \textit{position-extended Forest-Ruth-like} (PEFRL) scheme~\cite{Omelyan:CPC2002,Forest:PD1990} to ensure energy conservation. As in \citet{Cichowlas:PRL2005}, we take Taylor-Green vortex ($ k_0=1 $) as an initial condition, and  time evolve the system till 120 eddy turnover time ($ 2\pi/U_\mathrm{rms} $) with a constant $ dt=10^{-4} $. The total energy per unit volume, $ E =  \int d\textbf{r} (u^2/2)/\int d\textbf{r} = 0.125$, and it  is conserved up to 12 significant digits till the final time $ t=120 $. 	The system reaches equilibrium in approximately 100 eddy turnover time, consistent with the estimates of \citet{Cichowlas:PRL2005}  and \citet{Verma:arxiv2020_equilibrium}. In the asymptotic state, the velocity field is $ \delta $-correlated, and it is as random as that in a thermodynamic gas~\cite{Verma:arxiv2020_equilibrium}.

	\begin{figure}[hbtp]
		\includegraphics[width=0.8\linewidth]{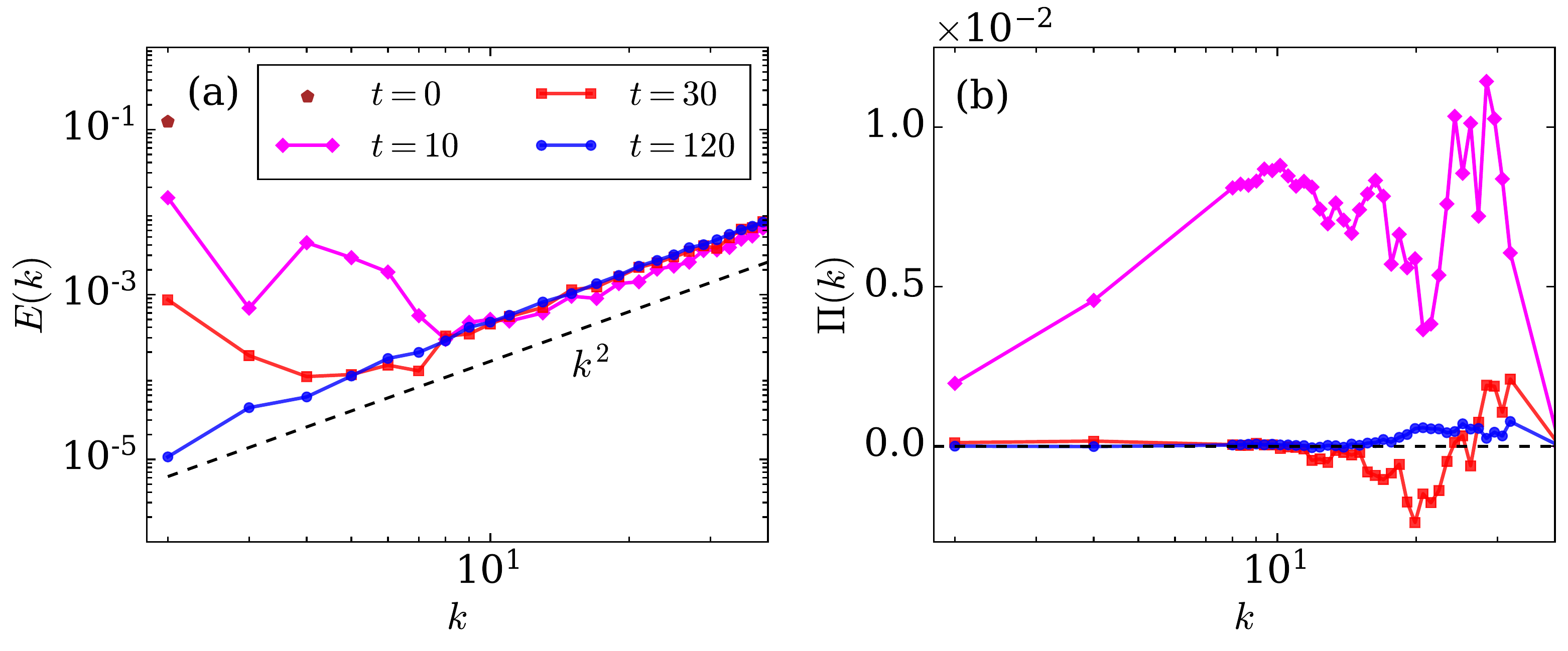}
		\caption{For 3D Euler turbulence, plots of  (a)  the energy spectra  $ E(k) $, and (b) the energy fluxes $ \Pi(k) $ at $ t=0,10, 30, 120 $. At $ t=120 $, $ E(k) \propto k^2 $.}
		\label{fig:3D_Ek}
	\end{figure}
Using the numerical data, we compute the  energy spectrum and flux for the flow at times $ t=0,10, 30, 120 $. For the computation of energy flux, we employ the algorithm outlined in \citet{Dar:PD2001} and \citet{Verma:PR2004}.  In Fig.~\ref{fig:3D_Ek}(a,b), we exhibit the energy spectra and energy fluxes at  $ t=0,10, 30, 120 $. Clearly, with time, the energy spectrum $ E(k) $  spreads from low wavenumbers to high wavenumbers,  and  asymptotes to $ k^2 $, consistent with the predictions of \citet{Kraichnan:JFM1973} [Eq.~(2)  with $ \gamma=0 $].

The energy flux is nonzero for the intermediate configurations, but for $ t \ge 30$, $ \Pi(k) \approx 0 $   for small $k $'s. Note, however, that $ \Pi(k) $ is of the order of $ 10^{-4} $ for large wavenumbers.  The fluctuations in the energy flux is suppressed significantly on averaging over 2000 frames in time interval (100,120) (see Fig.~\ref{fig:3D_Tk}).  We expect these fluctuations to subside at a later time when the system has thermalized fully.
\begin{figure}[hbtp]
	\includegraphics[width=0.8\linewidth]{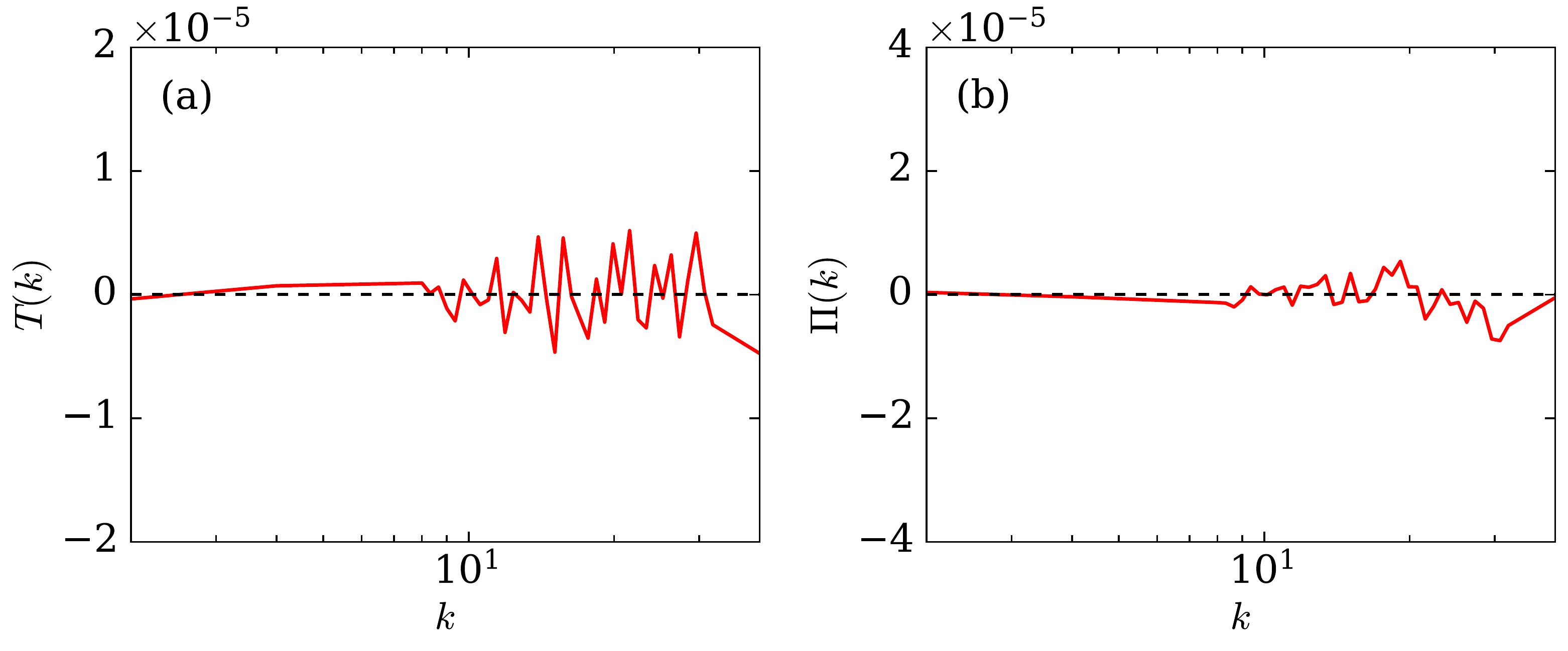}
	\caption{For 3D Euler turbulence, plots of (a) averaged     $ T(k,t) $, and (b) averaged $ \Pi(k,t) $.  We average over 2000 frames in time interval (100, 120).}
	\label{fig:3D_Tk}
\end{figure}}   

We also remark that  for  $ \delta $-correlated initial velocity profile, $ E(k) \propto k^2 $ [Eq.~(\ref{eq:Euler3D}) with $ \gamma=0 $], and  energy flux vanishes from the beginning itself~\cite{Verma:arxiv2020_equilibrium}. Thus, Euler turbulence remains thermalized throughout for $ \delta $-correlated initial condition.


%

\end{document}